# Tinkering in Primary School: From Episode to Science Practice

**Stefano Rini** – University of Bologna; Game Science Research Center, IMT School for Advanced Studies Lucca, IT

**Sara Ricciardi** – INAF Astrophysics and Space Science Observatory Bologna; University of Bologna; IAU Office of Astronomy for Education Center Italy; Game Science Research Center, IMT School for Advanced Studies Lucca, IT

### Abstract

This study discusses the opportunity to integrate tinkering, a constructionist practice, into formal education, highlighting its potential and challenges. We propose a model through which teachers can blend the open exploratory nature of tinkering with structured learning in primary school classrooms, focusing on Physics Education. Despite pandemic-induced limitations, feedback from 20 teachers and analysis of fishbowl protocols revealed the positive impact of tinkering on classroom dynamics, teacher engagement, and student access to knowledge. Our findings indicated that tinkering can surface relevant scientific questions. Nevertheless, teachers feel unprepared to tackle them in the classroom. This evidence will guide our future co-designs to enhance learning experiences and address the complexities of incorporating tinkering into formal education.

## 1. Tinkering: A Constructionist Practice for Full Scientific Citizenship

Tinkering is a constructivist practice traditionally rooted in informal settings, providing a holistic way to engage people in Science, Technology, Engineering and Mathematics (STEM) by blending it with art and combining it with high-tech and low-tech materials (Petrich et al., 2013; Resnick & Rosen-baum, 2013). It emphasises active knowledge construction, aligning with Papert's (1980) constructionism, which asserts that learning becomes meaningful when learners create personally significant artefacts. This method fosters creativity, exploration, and deeper understanding, particularly in self-directed projects.

Our team, composed of educators, teachers, and scientists, is committed to enhancing education through creativity, playfulness, and self-expression. Scientists in our group strive to demystify science, presenting it not as a collection of facts but as a dynamic, collaborative process filled with experimentation and discovery. Science's social nature – teamwork, idea-sharing, and collective refinement – is central to our approach. Some of our previous experiences are documented in Ricciardi, Rini, Villa, Ferrante et al. (2021) and Ricciardi, Rini, Villa (2021).

Tinkering practices can reveal the Nature of Science (NOS) (Lederman 1992, Erduran & Dagher 2014), promoting informed attitudes and critical engagement. Rooted in constructionist principles, tinkering mirrors scientific inquiry by blending creativity, collaboration, and active knowledge construction. Integrating these practices in public education is vital to cultivating democratic societies equipped to address complex challenges. We believe, in fact, that a practice like tinkering can, for all these reasons, serve as a concrete step toward a more inclusive and participatory democracy. This is because tinkering embodies a deep connection with science understood as a human, creative, and collaborative endeavour. Through tinkering, we can make the construction of scientific knowledge come alive from the bottom up, starting with tinkering experiences that can later evolve also through other tools and approaches. What matters is the consistent, personal and prolonged embodied connection with physical phenomena in a playful environment, which fosters authentic involvement. Allowing children's and students' research questions to emerge from their own explorations is, in our view, a crucial choice for nurturing critical thinking, scientific creativity, and citizenship. We view these aspects of our practice as critical for fostering scientific citizenship: empowering learners to participate actively in the knowledge society (Greco et al., 2008; Bandelli, 2016).

## 2. Constructionist Practice at Work: Coding and Tinkering

Italy's education system has embraced transitions inspired by the Lisbon Strategy (2000), emphasising STEAM approaches to promote interdisciplinary, creative learning. Coding initiatives, with Scratch as one of the most commonly used platforms, became very popular in 2015 to enhance digital literacy and computational thinking. However, as Resnick et al. (2020) highlight, coding's potential is undermined when approached rigidly, reducing it to rote tasks rather than tools for creative exploration. While coding gained traction, although frequently used in a limited way, tinkering remained marginalised, often limited to extracurricular settings. Its transformative potential lies in bridging disciplines and encouraging playful, exploratory learning. Yet, its lack of explicit curricular goals (Petrich et al., 2013; Bevan, Petrich et al., 2014) challenges its integration into formal education. The incredible power of tinkering lies not in disciplinary content but in how it enables us to understand the world around us. Through tinkering, one can learn and understand how we can construct knowledge as individuals and as a research community.

These ideas are most fully articulated in Papert's work, not only through his seminal publications (Papert, 1980, 1993), but also through the way he conducted his research, embedding his educational philosophy directly within school environments. Papert's constructionism radically reimagines learning/teaching as a process of active, creative engagement, in which knowledge is not transmitted but rather emerges through the construction of meaningful artefacts within socially and culturally rich contexts.

Emerging in the same historical period, these principles are closely aligned with the framework of critical pedagogy (Freire, 2018). Freire conceives of knowledge construction as a collective and dialogical process in which both teachers and students participate as co-investigators. This conception stands in stark opposition to the "banking model" of education, which Freire critiques as an oppressive approach wherein teachers "deposit" information into passive learners, denying them agency and critical thought. This shared legacy continues to shape contemporary educational practices that center learner agency, critical thinking, and creative exploration (Res-nick, 2017; Petrich et al., 2013; Bevan, Petrich et al., 2014; Martinez & Stager, 2019).

To tinker, observe, and reflect upon tinkering practices, we draw on educational approaches that place children's thinking and expression at the center. We are inspired by the Reggio Emilia tradition, which values children's multiple languages and emphasises documentation as a tool for interpretation and pedagogical reflection (Edwards et al., 2012; Giudici et al., 2011). Our perspective is also informed by the concept of playful learning, which highlights engagement, meaning, and joy as essential elements of deep learning (Zosh et al., 2018; Project Zero, 2016).

It is within this framework that we situate our research, exploring how tinkering, as both a pedagogical stance and a design practice, can support the development of epistemic curiosity and foster authentic engagement with scientific phenomena in primary education, contributing to more inclusive and democratically grounded science learning environments.

## 3. Officina della Luce (Light Workshop): Tinkering in the Classroom

Since 2012, we have worked with teachers to introduce tinkering in schools through workshops with students and co-design processes with educators. Our experiments were successful, and teachers appreciated and used constructivist practices, but they were often relegated to recreational time or time for "other activities", not learning itself. Over the years, we have conducted and recorded interviews with teachers to better understand and re-orient our work, which we are still analysing. The first layer of the analysis suggests that tinkering was perceived as something interesting and exciting, sometimes transformative, but not directly interacting with school life. The crucial issue is that teachers often struggle to fully unpack all the physics embedded in a tinkering workshop. Since tinkering is a creative interaction with a physical phenomenon, the teacher-facilitator should have a deep and nuanced understanding of that phenomenon in order to recognise and support the cognitive challenges students are facing at any moment. They should also be able to connect emerging questions with other experiences that can deep-en and extend the investigation. We've observed that teachers tend to use tinkering in ways that feel more familiar to them – for example, using light play as a storytelling exercise rather than as an opportunity for deep inquiry into the phenomenon of light. This is not because students lack curiosity, but because the teacher feels more confident staying in a known territory rather than venturing into the unknown.

We realised that achieving a deeper and more significant impact required meeting teachers' expectations and unpacking those workshops. Through co-design discussions, it became evident that teachers considered having a clear and structured link with learning objectives crucial. This need is legitimate for teachers, but it puts us in a dilemma because forcibly attaching a disciplinary goal to tinkering would completely distort this practice, reducing its essence significantly. We did not want these practices to lose their potential, as was sometimes happening with coding being reduced to its engineering and technical side. We did not solve this problem quickly; some ideas began to form by delving into pedagogical activism, especially in the works of Malaguzzi, Lodi, and Ciari, and also by studying Rodari and Munari. Eventually, Ciari's text, "The New Educational Techniques", clarified our thoughts and led us to the model we attempted to implement in our "Officina della Luce".

This also ties into our particular interest in the Sciences. We believe that the core of scientific learning lies in the understanding of the Nature of Science (NoS). Recognising this dimension as foundational not only enables a deeper grasp of the discipline's meaning – particularly in the case of physics – but also represents an essential step toward the development of a conscious and democratic citizenship. When this perspective is adopted, many subsequent educational choices become clearer and more coherent: teachers may be more inclined to reconsider the centrality of disciplinary content in favour of deeper cognitive and emotional engagement, aimed at fostering a meaningful and authentic relationship with physics and the sciences more broadly.

In our view, promoting a genuine understanding of the nature of physics requires an experiential approach: physics must be lived, reconstructed through educational pathways that reflect its epistemological foundations – that is, the authentic processes through which scientific knowledge is constructed and evolves. A curriculum designed in this way, grounded in ex-ploration, foundation of a research question, argumentation, hypothesis formulation, and testing, proves essential in making learning truly meaningful.

Nevertheless, many active learning practices fall short of their potential precisely because they overlook what we consider to be the first and most crucial step in the process of scientific inquiry: the formulation of one's own research question. Without this generative element, school experiences the risk of becoming a sequence of activities devoid of real epistemic value for the learner. In this regard, we argue that tinkering— an open-ended, creative practice—offers a privileged context for the emergence of authentic student questions. Once made explicit, these questions can be nurtured and developed through a variety of learning experiences, contributing to the construction of scientific knowledge that is both personal and shared.

Tinkering could be a fundamental practice and technique for the science class: a tinkering workshop can be seen as a playroom where children can find their relevant research questions while personally engaging with different phenomena. Learners allow themselves to fail, collaborate, exchange knowledge, explore materials, make hypotheses, and test their theories, working the same way a scientific community works. When tinkering happens in a classroom, common knowledge emerges even if different groups tackle different problems with different ideas. This core of personal and significant but necessarily incomplete knowledge can be a stimulus for other tinkering sessions, but also can lead to different experiments and explorations that the classroom can plan together. Of course, these further explorations will have specific learning objectives and could be integrated into the curricula. Tinkering could be a precious open moment where students get to experiment creatively with the world around them by observing and understanding, building artefacts, and constructing their microworlds. During this open exploration, questions may arise, and the class, as a learning community, can collectively try to answer these, building a contraption or setting up an experiment to do so. The class members can also search for information in the school library or online, or in external resources. They can engage with different people from the larger community outside the school with a specific professional or academic background that resonates with the new open questions. A self-posed question is invaluable for children's agency and learning.

It will nurture their curiosity and passion and help build their confidence to tackle complex problems.

Our collaborative team, comprising researchers and educators, endeavoured to conceptualise an educational pathway that synergises tinkering workshops with the methodologies commonly employed in school contexts, from hands-on experiments to books, textbooks, and audiovisual resources. The aim was to craft a learning journey favourable to the collective knowledge construction by the classroom community.

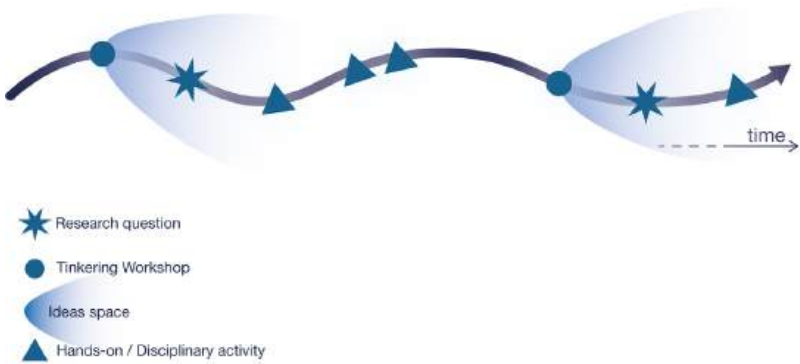

Figure 1 – The overall scheme of TIDE, a possible integration of Tinkering with disciplines.

So, with this idea in mind, we composed TIDE (Tinkering, Ideas generation, Disciplinary connection, Exploration), a preliminary and simple model that combines tinkering with disciplinary learning as in Figure 1. The wavy line represents a timeline that describes what happens in the classroom; this is a very simplified image given that it tries to depict the complex life of a classroom, and it represents a shift from the even simpler trajectory of the traditional classroom where a more instructionist approach (Papert, 1993) is in place and where the classroom's life is reduced to a series of juxtaposed and predefined learning experiences and goals.

We start with - or propose at some point - a tinkering workshop that, with its open, playful and exploratory nature, triggers various questions among the students, some of which may be answered during the workshop, but many remain open for further exploration. Each tinkering workshop is cen-tred on a specific material or physical phenomenon, leading us to believe that the resulting idea space will be rich but relatively compact. The cloud of ideas represents different possible explorations generated from the tinkering experience. This state creates a fertile ground for connecting with disciplines. As Ciari states, a sensitive teacher will know how to choose. Being closest to the students, the teacher will discard unsuitable themes and delve deeper into the most significant and urgent questions and issues that align with the children's development. At this point, teachers can work with students using all the tools in their repertoire, books, illustrations, and other media, conducting experiments, and returning to tinkering with different skills and perspectives. During these subsequent moments, after a specific research question has surfaced throughout and after the tinkering sessions, the learning objective is evident to the teacher and the learners. Together, pupils and teachers will engage in re-constructing a piece of knowledge.

The only risk of this approach is that the two phases become blurred. Tinkering must remain an open-ended experience; however, this openness could lead to questions that teachers might feel unprepared to answer. Embracing children's questions means opening an investigation and accepting a temporary inadequacy. This process could be intimidating, but it is precisely the techniques of tinkering that can help. Tinkering practices force teachers to rethink their way of being in the classroom profoundly. To facilitate tinkering, the teachers must step out or step aside from their traditional role, constantly refocus on what is relevant and meaningful to the students, and allow learning to be constructed together through the classroom's learning practice. Tinkering empowers teachers, equipping them with the skills to think together through things and to guide exploration even in the absence of profound knowledge of the answers. Maintaining the facilitator's stance even during investigations related to disciplinary objectives is crucial for a real co-construction of learning.

The most ambitious goal of our project was to understand if, with appropriate tools and support, teachers could integrate tinkering practices deeply into classroom life. This integration could happen at different levels, depending mainly on the classroom's general conditions and the teachers' willingness to work in this direction.

## 4. Officina Design and Its Actual Implementation

We now describe a specific project in which, for the first time in a structured manner, we attempted to implement the TIDE model, particularly an experimental project from September 2021 to September 2022 that involved 13 primary school classrooms and 24 teachers in Bologna. In preparing for this project, the leading school won a grant. The dedicated fund covered materials, teachers' extra time, including documentation time, and external expert facilitators to sustain the teachers' actions.

Our goal was to determine whether, given access to multiple tinkering sessions, a repository of possible disciplinary connections, and documentation tools, teachers would deeply integrate tinkering into school life. We did not structure a detailed work plan for the teachers as we believed it was important to understand whether they could independently design an educational project centred on tinkering. We provided professional development sessions where participants could personally experience all the tinkering workshops later implemented in the classroom. During the three full training days, we also introduced the available resource library and tested several documentation tools presented by INDIRE (Istituto Nazionale di Documentazione, Innovazione e Ricerca Educativa) researchers. Besides designing the documentation structure, they facilitated some sessions devoted to documentation.

The central theme of this project is light. We selected the Light Play workshop as the main tinkering activity proposed at least three times for each class. We recontextualised the original workshop designed by the Tinkering Studio by modifying some of the materials, the setting, and the facilitation. We also developed a library of educational resources around this theme. We organised the material into three branches, documented on the INAF online platform https://play.inaf.it/officinadellaluce/
- Tinkering workshops: workshop description, possible facilitation, materials and possible bridges with other workshops and art.
- Reference materials and hands-on science: a mini library of hands-on activities with explanations from which the teacher can draw a lesson plan or teaching ideas. This work was prepared by INAF researchers from the Creative Learning, Tinkering, and Games working group, who developed and published this online repertoire.
- Documentation: Besides informal communication, we set up a documentation protocol that teachers could use to communicate with each other and with us: the fishbowl protocol. It is a reflective technique developed by Project Zero, a device aimed at building a safe and welcoming environment for teachers to discuss a piece of documentation collected from a learning activity. It is built on a few precisely timed steps in which one of the teachers shows a piece of evidence to reflect upon and relaunch their action in the classroom (Giudici et al. 2001, Mughini et al. 2020).

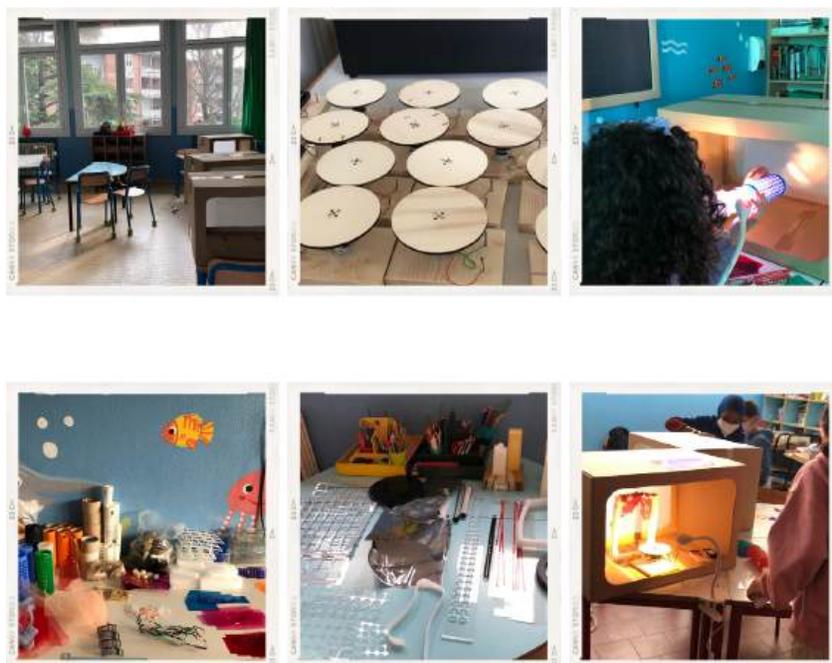

Figure 2 – The school setting, materials and students working in the Light Play workshop

In this first experiment, we did not provide teachers with a predefined path. Instead, we offered methodological training on tinkering, and we provided a digital library of hands-on workshops and educational resources that teachers could freely use to integrate tinkering into their disciplinary teaching as they saw fit. We also offered teachers the opportunity to involve us directly—either to lead one of the proposed workshops in the classroom or to support other activities they wanted to offer their students but didn't feel confident facilitating on their own. Three classes invited us to collaborate in experimenting with additive and subtractive colour mixing.

The initial classroom design included five tinkering workshops interspersed with documentation and documentation-sharing activities between teachers and researchers starting in September 2022. Four primary schools were involved in the area of Bologna, with 13 classrooms and 24 teachers (approximately 300 students). Every classroom experiences tinkering workshops at least five times (10 hours min.). Unfortunately, the evolution of the pandemic forced us to compress the project into four months instead of the planned eight months. Thus, we modified the original design for organisational reasons and to address the needs of the pupils still in a problematic situation. In fact, during the post-pandemic year, we had to re-focus on cooperation and playful interaction with peers because teachers reported that students had almost lost these essential skills. Children continuously asked for permission to touch materials, share them, and interact with each other.

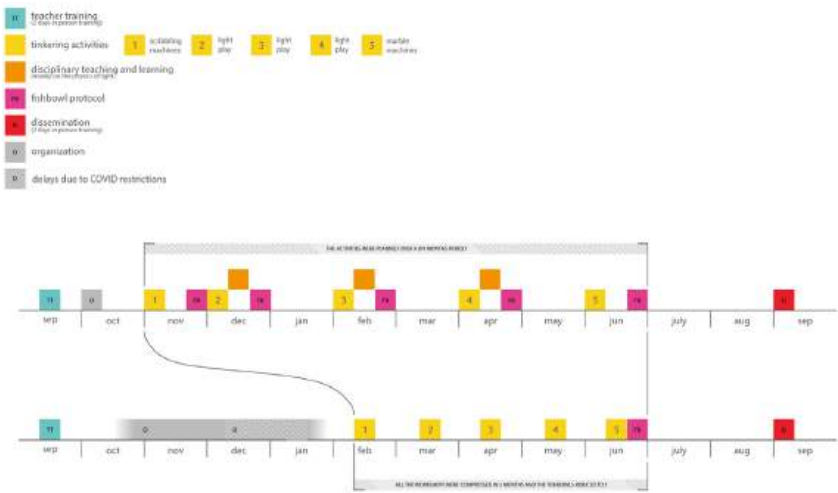

Figure 3 – Officina della Luce: planned and actual project activity schedule

We refocused our research together with teachers, allowing them to reorganise the activities and autonomously plan individual projects while still supporting the planned ones. Unfortunately, this problem significantly impacted documentation because it was only possible to run the fishbowl documentation protocol once for each team of teachers and only at the end of the program, after they had done all the workshops. So, while the fishbowl moments were initially intended as a time for sharing within the group of teachers, for reflecting and subsequently relaunching action in the classroom, they became more of a final reflection moment.

Documentation was carried out by the classroom teachers themselves. Each tinkering session typically involved one teacher acting as co-facilitator and another as documenter, while one or two external facilitators led the activity. The documentation formats included written notes, photographs, and short video recordings, some of which were shared with families or transformed into class product (https://tinyurl.com/tinkeringIC11). As researchers, we only accessed the materials that teachers chose to share—either during the fishbowl discussions or through informal exchanges between workshops. Due to pandemic-related constraints, the documentation component was inevitably the most impacted and remained the most fragmented part of the project. Despite these challenges, we can draw conclusions that are already helping us design the next steps.

At the end of the activities with the pupils, we designed a questionnaire proposed in September 2023 during the project's feedback session with all participants. The questionnaire aimed, on the one hand, to quantify how capable teachers felt in integrating tinkering with learning objectives despite the challenging context; on the other hand, we took the opportunity to measure and delve deeper into what many teachers had frequently reported over time: the unexpectedly high engagement, participation, and effectiveness observed in students often regarded as disengaged from school or even problematic. In the following paragraphs, we examine this questionnaire completed by 20 teachers in the program. We will also discuss the analysis of a recorded fishbowl protocol that will shed light on teachers' attitudes regarding science teaching and their possible uneasiness in accepting and relaunching scientific questions when they do not feel prepared enough.

## 5. "School-Oriented" and "Non-Aligned" Students in the Tinkering Workshop

Through this analysis, we aim to investigate how different types of students may benefit—or somehow fail to benefit—from tinkering practices. This research question emerged because, over the years, teachers reported that the students with whom they had the least expectations were often the ones who performed well in tinkering activities and were sometimes even activity leaders. At the same time, some students whose teachers expected them to perform very well encountered significant difficulties in tackling the workshop. In particular, an otherwise highly performing student almost refused to participate and said, "I'm not doing the activity; I'm just helping them", shielding themselves from the possibility of failure.

In the questionnaire, we asked if the tinkering practice revealed unexpected or partially expected behaviour, and then we asked them to express what they noticed. 70% of the teachers answered positively, commenting, "Children who struggle the most with traditional educational activities showed they could navigate them easily and enthusiastically; a very academically proficient girl, on the other hand, experienced several frustrations. Everyone showed enthusiasm when working in pairs/groups, even with classmates with whom they often conflict." "Children who usually do not take on a leading role during traditional lessons became protagonists within the small group, while, on the contrary, some children who are usually considered 'capable' felt unsettled by the practical task."

Through two additional specific questions, we asked:
1. "How did the highly "school-oriented" children (those who naturally fit into the school system) perform during the Tinkering workshops?"
2. "How did the less "school-oriented" children (those who show little interest during school activities or struggle to adapt to the mechanisms of school) perform during the Tinkering workshops?"

From now on, we will refer to these two interpretative clusters as "aligned" students and "non-aligned" students, based on patterns that emerged inductively from teachers' feedback during post-workshop discussions. These last kids are the ones who more often have problems finding their motivation in school and accepting the proposed activities. We analysed the answers of 20 teachers who participated in the study, coding the teachers' brief descriptive texts using seven themes for "aligned" students and eight themes for "non-aligned" students, as illustrated in Figures 4 and 5.

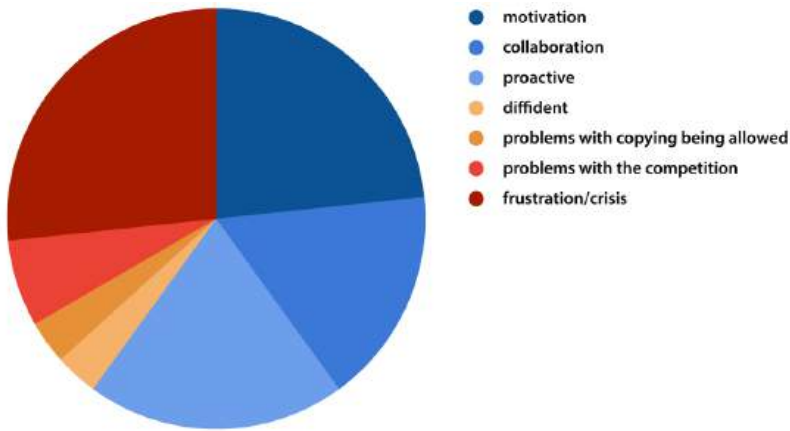

Figure 4 – Coded answer to question 1: aligned students

As evident from the chart in Figure 4, some students from the aligned group respond very positively to tinkering, while for others, it becomes a source of frustration and, at times, even crises.

The majority of responses, although positive, highlighted the presence of factors related to frustration and crises. For instance, one states, "School-oriented children learned to manage the frustration of not achieving immediate success. They became more relaxed, focusing more on the process and reasoning. They reconciled with the possibility of making mistakes." "The school-oriented children greatly appreciated the importance and value of collaboration, self-reflection, and mutual assistance. Some were 'challenged' by the possibility of being allowed to copy, while others found the competitive nature of the marble machine workshop to be a source of difficulty."

We conducted the same analysis for the second question dedicated to "non-aligned students". Also, in this case, each text was eventually associated with more than one theme. What is evident is that teachers report for non-aligned students extremely positive attitudes within the tinkering practices.

We report some representative answers to get the general tone: "They participated willingly and even felt like protagonists." "They responded positively by actively participating, revealing their skills and richness." "High motivation, sustained attention and engagement over time, and creativity.".

We also report the two answers coded as non-completely positive: "Some of them demonstrated a greater predisposition, while others maintained a 'delegation' attitude toward the rest of the group. To address this, we continuously adjusted the group configurations to encourage an active and participatory attitude from everyone". "Some performed well. In contrast, others (one in particular) were easily distracted and struggled to pursue a specific goal."

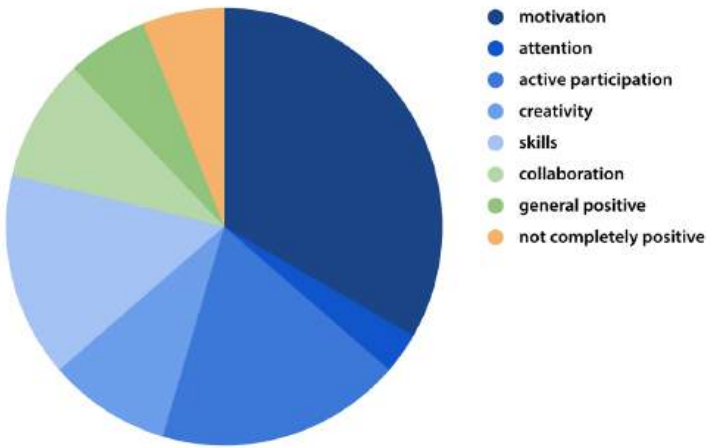

Figure 5 – Coded answer to question 2: non-aligned students

In the end, we discussed these graphs with the teachers, identifying at least three student types based on these observations: The first type describes students who grasp the teacher's requests and suggestions, understand the school dynamics, but can also focus on their interests, being able to develop them within the school context. These are the aligned students with a positive experience during tinkering, represented in Figure 4 from light to dark blue sector. The second type is a student who understands the teacher's requests and suggestions, is aware of the school dynamics, and is highly focused on meeting these demands rather than pursuing their own interests.

These are the aligned students with a challenging experience during tinkering, represented in Figure 4, orange to red sector. The third type is a student who is either unable or uninterested in responding to the teacher's requests and suggestions, does not grasp the school dynamics, and is intensely focused on their own interests. These are the non-aligned students, represented in Figure 5. This classification depends on the student's school experience, personal history, and situation.

The fact that students considered generally uninterested respond exceptionally positively to these activities has led teachers to reflect deeply. They have seen firsthand that we can genuinely engage students on the school path's margins with interventions like tinkering. Similarly, revealing the fragility of students whose motivation in school activities relies almost entirely on mutual recognition between teacher and student is significant, as these students—often seen as high achievers—are, in reality, quite vulnerable.

The teachers discovered a powerful tool for re-engaging students who are often passive or disinterested. At the same time, they became aware of vulnerabilities that had not surfaced within the school routine. In this sense, the Tinkering workshop was helpful to observe the students in an unusual context, allowing for a deeper understanding of them.

This first analysis helped us refine our focus and better understand what we should investigate further. While the findings offer valuable insights into student behaviours, they also reveal important aspects of the school system in which these students operate, as well as the underlying assumptions and pedagogical orientations of their teachers. Our broader field observations—beyond this specific experience—suggest that teachers who expressed the greatest surprise at students' agency during tinkering activities were often those more accustomed to formal, content-driven instructional models, where creativity and self-expression are not central. Conversely, in classrooms where individual contributions and expressive approaches are regularly valued, such reactions were less pronounced or absent.

This discrepancy points to the need for a deeper exploration—not only of student profiles, but also of how different teaching styles influence the perception and development of student agency in open-ended contexts like tinkering. Further research should include more targeted analyses of student trajectories, ideally cross-referenced with teachers' pedagogical beliefs and classroom practices.

# 6. Physics: Students' Research Questions and Teachers' Comfort Zone

Some interesting results and themes emerged from the preliminary analysis of some fishbowl protocols. Specifically, we report here the study of a fishbowl protocol where the documenting teacher teaches humanities, and thus, integration with the science curriculum was not initially planned. For the fishbowl, the teacher selected a fragment of documentation where the students enthusiastically raised a relevant research question related to the physics of light. They wondered how to produce white light using the available light bulb (white) and various coloured plastic filters. Initially raised by just one group, this question spread throughout the workshop, eventually challenging the entire class.

She commented on the documentation and reported that the students tried to apply prior knowledge. Remembering the phrase "the sum of all colours makes white", they attempted to overlap all the gel sheets to achieve white light but without success because when all filters are combined, they block all the light, which ultimately does not pass through. The teacher admitted to us that at that moment, she felt challenged because she did not know how to answer the students' questions. As a result, the significant research question was left unresolved.

As the documentation session progressed, the teacher realised what had happened and recognised her difficulty in fully understanding and addressing the students' profound research questions. While tinkering, the learners wondered why what they had learned the previous year, studying the eye (additive mixing), did not apply in this case (subtractive mixing). The documenter/teacher recognised that it would have been essential to address and explore this crucial question, but also acknowledged, along with the other teachers, that they were not ready from a disciplinary standpoint.

The vital evidence of this documentation is twofold: Tinkering sparked an essential and profound research question, and the teacher recognised her inadequacy in embracing this question. We noticed that deep experimentation with light and matter raises many theoretical and abstract questions as the previous. Similarly, many questions arose because of the desire to achieve a specific aesthetic or narrative result, such as: "I want to create a marine back-drop and have lights going in all directions that look like sparkles; how do I do that?" This question often allows learners to reflect and wonder about another disciplinary idea of the physics of light: reflection. Here, we observe what we are accustomed to seeing in many tinkering sessions, as also investigated in Bevan, Gutwill et al. (2014): many research questions emerge and are partially resolved during tinkering sessions.

Let us now focus on the teachers' sense of inadequacy, particularly regarding the physics of light. At the end of the documentation session, the teacher expressed that it would be necessary to integrate some form of disciplinary training to help teachers feel more competent and, consequently, better equipped to guide students' inquiries. This same sense of inadequacy or uneasiness is also evident in the teachers' preferences when selecting a disciplinary area to investigate further after the tinkering activity. After being trained in tinkering practices and following the presentation of resources on light and documentation, teachers were free to decide independently if and how to make curricular connections, which were entirely optional and voluntary.

At the end of the project, we aimed to gain a comprehensive understanding of the disciplinary areas autonomously explored by the teachers. So we included two questions in the final questionnaire for the teachers: "Have you carried out activities connected to or stemming from the tinkering workshop?" and then "In activities not conducted by external experts, in which area did you primarily work with your class?"

We organised the responses of the 20 teachers into a chart that highlights their preferences. The first significant finding is that the majority of teachers did not isolate tinkering activities but chose to integrate the workshops into the life of the classroom, albeit through pathways of varying intensity and complexity.

Despite the availble educational library was centered on science curricula, the teachers vastly preferred integrating activities focusing on language, expression, and storytelling. This shift may have been accentuated by the particular historical context experienced and the need to reconnect with personal expression, but also by the particular tinkering workshop proposed in fact creating kinetic light sculptures may easily suggest a storytelling approach.

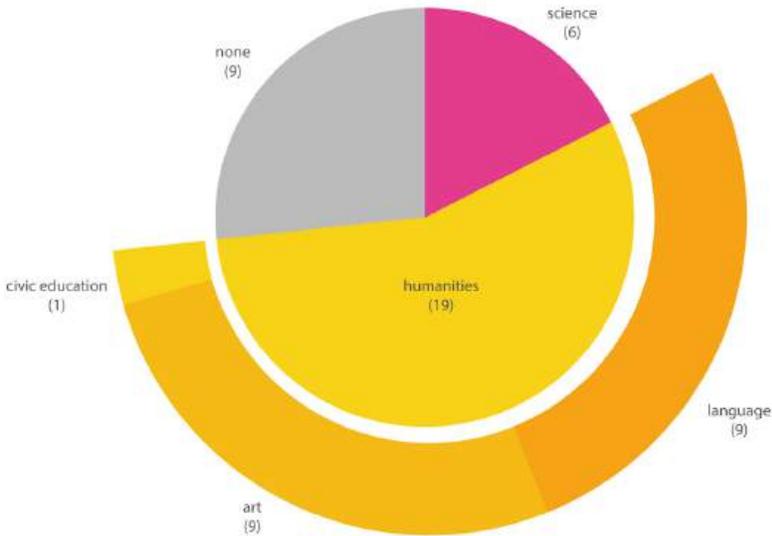

Figure 6 – "Integrating tinkering with disciplinary learning in which area did you primarily work?"

To provide just a few examples: some teachers developed a complete PBL (Project-Based Learning) project, creating a collective storytelling experience and subsequently presenting it to their peers, embarking on a complex and extended activity. Another class produced a video by working with light play and creating a complementary sound design, drawing on the students' diverse knowledge and emphasising teamwork throughout the process. Yet another class focused primarily on disciplinary aspects related to physics, conducting a series of experiments and hands-on activities.

## 7. Conclusion and Future Perspectives

The research path presented here is the result of approximately three years of work within schools, aimed at observing how tinkering and constructionist practices influence teachers' daily routines.

For the first time, we introduce TIDE, a potential approach to integrating tinkering with learning objectives. While situated in a specific context, our research allows us—albeit preliminarily—to draw meaningful conclusions and observations that are valuable for assessing the progress made so far and reorienting our stance and research questions for future developments:

A key finding is teachers' uneasiness in engaging with scientific disciplinary exploration. Teachers often feel insufficiently competent in scientific subjects, particularly physics. This is evident from the responses in Figure 6, which summarise the teachers' preferences toward humanities or expressive domains for integrating tinkering into the curriculum. The analysis of the fishbowl conducted in the previous paragraph aligns with this direction: even when an honest and authentic scientific question emerged in the classroom, teachers did not feel equipped to address it or adopt the research question as the focus for subsequent activities.

Tinkering in schools has proven to be a powerful tool for re-engaging students who often struggle with traditional educational mechanisms while simultaneously revealing potential vulnerabilities among some high-performing students. On one hand, this shifts teachers' perception of their students; on the other, it provides an opportunity for teachers to transform their attitudes toward learning processes.

Although not implemented as initially planned, the documentation process, particularly the fishbowl protocol, effectively highlights fundamental aspects of the ongoing educational process that might otherwise have remained invisible. We believe that documentation may create opportunities for further intentional relaunch.

These conclusions indicate that, in continuing this action-research process, it will also be essential to provide teachers with support in disciplinary preparation. The training and co-design phases, which until the Officina della Luce focused on the methodological aspects of tinkering and documentation practices, should also include dedicated sessions addressing appropriate disciplinary content.

To fully validate the TIDE model, it will be necessary to document children's ideas over time, exploring their learning outcomes and the development of scientific thinking that frames knowledge as a field of inquiry and experimentation rather than as a set of crystallised facts.

### Acknowledgements

The authors would like to sincerely thank the anonymous reviewers, whose thoughtful feedback and encouragement made us believe that it was worthwhile to share not only the practices we developed but also the deeper ideas and reflections that shaped them.